\documentclass[]{spie}  

\usepackage{amsmath,amsfonts,amssymb}
\usepackage{graphicx}
\usepackage[colorlinks=true, allcolors=blue]{hyperref}
\usepackage{float}
\usepackage{pifont}

\title{Development status of the UV-VIS detector system of SOXS for the ESO-NTT telescope}

\author[a,b]{Rosario Cosentino}
\author[a]{Marcos Hernandez}
\author[a]{Hector Ventura}
\author[c]{Sergio Campana}
\author[d]{Riccardo Claudi}		
\author[e]{Pietro Schipani}	
\author[c]{Matteo Aliverti}		
\author[d]{Andrea Baruffolo}	
\author[f,g]{Sagi Ben-Ami}		
\author[h]{Federico Biondi}		
\author[e]{Giulio Capasso}		
\author[i]{Francesco D'Alessio}
\author[c]{Paolo D'Avanzo}		
\author[f]{Ofir	Hershko}		
\author[j,k]{Hanindyo Kuncarayakti}
\author[c]{Marco Landoni}			
\author[b]{Matteo Munari}			
\author[l,m]{Giuliano Pignata}		
\author[n]{Adam Rubin}				
\author[b,o]{Salvatore Scuderi}		
\author[i]{Fabrizio Vitali}			
\author[p]{David Young}				
\author[q]{Jani	Achrén}				
\author[l,r]{José Antonio Araiza-Duran}
\author[s]{Iair	Arcavi}			
\author[m,t]{Anna Brucalassi}	
\author[f]{Rachel Bruch}		
\author[d]{Enrico Cappellaro}	
\author[e]{Mirko Colapietro}	
\author[e]{Massimo Della Valle}	
\author[d]{Marco De Pascale}	
\author[b]{Rosario Di Benedetto}
\author[e]{Sergio D'Orsi}		
\author[g]{Avishay Gal-Yam}		
\author[c]{Matteo Genoni}		
\author[j,k]{Jari Kotilainen}	
\author[u]{Gianluca Li Causi}	
\author[j]{Seppo Mattila}		
\author[g]{Michael Rappaport}	
\author[d]{Kalyan Radhakrishnan}
\author[d]{Davide Ricci}		
\author[c]{Marco Riva}			
\author[d]{Bernardo Salasnich}
\author[p]{Stephen Smartt}
\author[b]{Ricardo Zanmar Sanchez}
\author[v]{Maximilian Stritzinger}
\author[n]{Matteo Accardo}
\author[n]{Leander H. Mehrgan}
\author[n]{Derek Ives}
\author[n]{Josh Hopgood}

\affil[a]{INAF - Fundaci\'{o}n Galileo Galilei, Bre\~{n}a Baja, Spain}
\affil[b]{INAF - Osservatorio Astrofisico di Catania, Catania, Italy}
\affil[c]{INAF - Osservatorio Astronomico di Brera, Merate, Italy}
\affil[d]{INAF - Osservatorio Astronomico di Padova, Padua, Italy}
\affil[e]{INAF - Osservatorio Astronomico di Capodimonte, Naples, Italy}
\affil[f]{Weizmann Institute of Science, Rehovot, Israel} 
\affil[g]{Harvard-Smithsonian Center for Astrophysics, Cambridge, USA}
\affil[h]{Max-Planck-Institut für Extraterrestrische Physik, Garching, Germany}
\affil[i]{INAF - Osservatorio Astronomico di Roma, Rome, Italy}
\affil[j]{Tuorla Observatory, Department of Physics and Astronomy, University of Turku, Turku, Finland}
\affil[k]{FINCA - Finnish Centre for Astronomy with ESO, Turku, Finland}
\affil[l]{Millennium Institute of Astrophysics (MAS), Santiago, Chile}
\affil[m]{Universidad Andres Bello, Santiago, Chile}
\affil[n]{European Southern Observatory, Garching, Germany}
\affil[o]{INAF - Istituto di Astrofisica Spaziale e Fisica Cosmica, Milano, Italy}
\affil[p]{Queen's University Belfast, Belfast, UK}
\affil[q]{Incident Angle Oy, Turku, Finland}
\affil[r]{Centro de Investigaciones en Optica A. C., León, Mexico}
\affil[s]{Tel Aviv University, Tel Aviv, Israel}
\affil[t]{INAF-Osservatorio Astrofisico Arcetri, Firenze, Italy}
\affil[u]{INAF - Istituto di Astrofisica e Planetologia Spaziali, Rome , Italy}
\affil[v]{Aarhus University, Aarhus, Denmark}
\authorinfo{Further author information: (E-mail: cosentino@tng.iac.es, Telephone: +34 922433642\\ }
 
\begin{document} 
\maketitle

\clearpage
\begin{abstract}
SOXS will be the new spectroscopic facility for the ESO NTT telescope able to cover the optical and NIR bands by using two diﬀerent arms: the UV-VIS (350-850 nm), and the NIR (800-2000 nm).
In this article, we describe the development status of the visible camera cryostat, the architecture of the acquisition system and the progress in the electronic design. 
The UV-VIS detector system is based on a CCD detector 44-82 from e2v, a custom detector head, coupled with the ESO continuous ﬂow cryostats (CFC), a custom cooling system, based on a Programmable Logic Controller (PLC), and the New General Controller (NGC) developed by ESO. This paper outlines the development status of the system, describes the design of the diﬀerent parts that make up the UV-VIS arm and is accompanied by a series of information describing the SOXS design solutions in the mechanics and in the electronics parts. The first tests of the detector system with the UV-VIS camera will be shown.

\end{abstract}

 (Ref. ~\citenum{soxsold,soxscosentino,soxsaliverti,soxsbiondi,soxsbrucalassi,soxscapasso,soxsclaudi,soxssanchez,soxsschipani,soxsricci,soxsrubin,soxsvitali,ngcpaper,aliverti,biondi,brucalassi,claudi,colapietro,genoni,kunca,landoni,ricci,rubin,sanchez,scuderi,shipani,vitali,young}).




\section{The UV-VIS detector system}
The UV-VIS CCD Detector System for SOXS is based on the e2v CCD44-82, the ESO NGC CCD controller unit and an ad hoc designed camera.

\subsection{The Detector}
The e2v detector is a back illuminated CCD with a 15 $\mu$  square pixel and an image area of 30.7 x 61.4 mm. 
The high quantum efficiency (QE) in the spectral response of the spectrograph (350-850 nm) and their geometric characteristics (pixel size and dimensions) make this CCD the most suitable for our instrument.

 \begin{figure} [H]
   \begin{center}
   \begin{tabular}{c} 
   \includegraphics[height=6 cm]{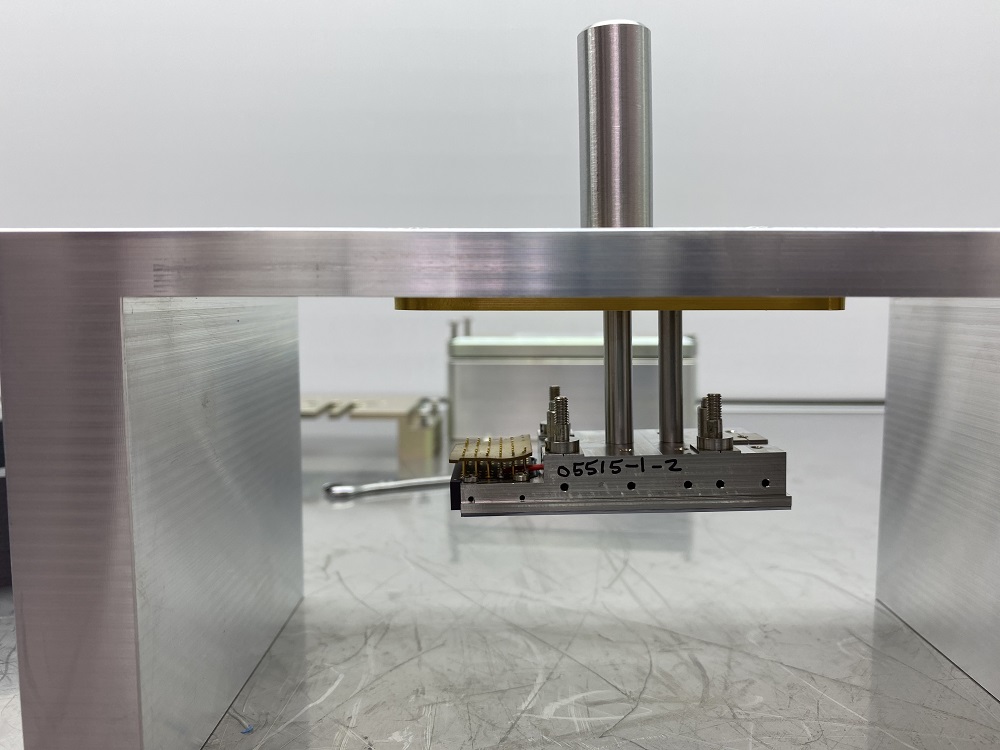}
   \end{tabular}
   \end{center}
   \caption[VIS Camera] 
   { \label{fig1} The engineering CCD on the mounting facilities hardware.}
   \end{figure}

\subsection{The NGC controller}

The ESO CCD controller (NGC) is a programmable electronic system for the UV-VIS CCD Detector polarization and readout. The flexibility in the programming of the operative parameters (bias and clock voltages, clock sequences, electronic gain) of the CCD and the features provided for the testing of the signals (programmable test points on the NGC front panel) helps the realization of the test and the optimization of the CCD operative conditions. 
In addition, the NGC is fully compatible with ESO software and it makes it easier the integration of the SOXS spectrograph in the software NTT environment.

   \begin{figure} [H]
   \begin{center}
   \begin{tabular}{c} 
   \includegraphics[height=8.1 cm]{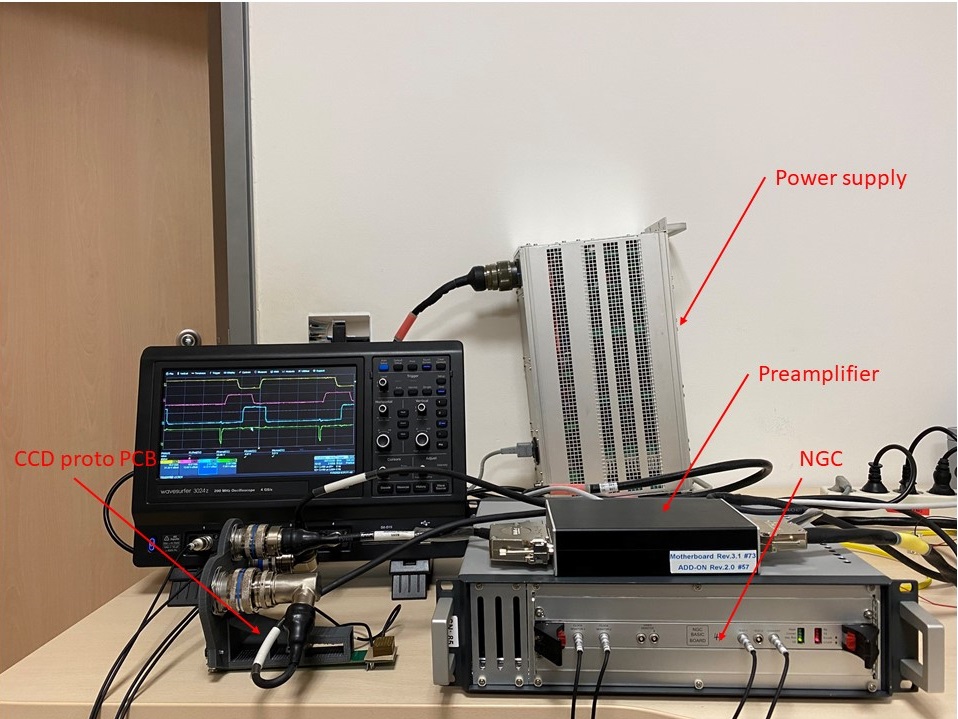}
   \end{tabular}
   \end{center}
   \caption[VIS Camera] 
   { \label{fig2} The CCD test bench with the NGC and test facilities.}
   \end{figure}

\subsection{The camera housing}

The camera is made up by the cryogenic part, based on the Continuous Flow Cryostat (CFC) coupled with the UV-VIS camera and the wiring for the detector and for the temperature sensors and control. The mechanic design of the camera is optimized to be mounted inside the reduced dimension of the UV-VIS spectrograph with the consequence of a reduced space inside the camera for the electronic connections. 
The reduced space inside the camera led to the development of an optimized Printed Citcuit Board (PCB), based on flex technology, for the connection of the CCD with the front-end electronics (Figure ~\ref{fig4}).

   \begin{figure} [H]
   \begin{center}
   \begin{tabular}{c} 
   \includegraphics[height=8.5 cm]{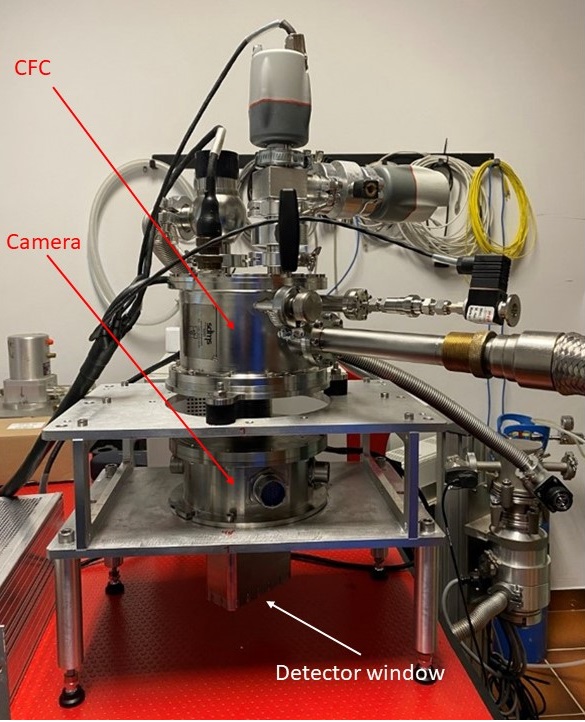}
   \end{tabular}
   \end{center}
   \caption[VIS Camera] 
   { \label{fig3} UV-VIS camera coupled with CFC.}
   \end{figure}

\subsection{The internal wiring}

For the internal wiring we adopted a solution based on a flex cable for the CCD connection, a coaxial cable for the video signals and a twisted vacuum wires for the temperature reading and control.
   \begin{figure} [H]
   \begin{center}
   \begin{tabular}{c} 
   \includegraphics[height=8 cm]{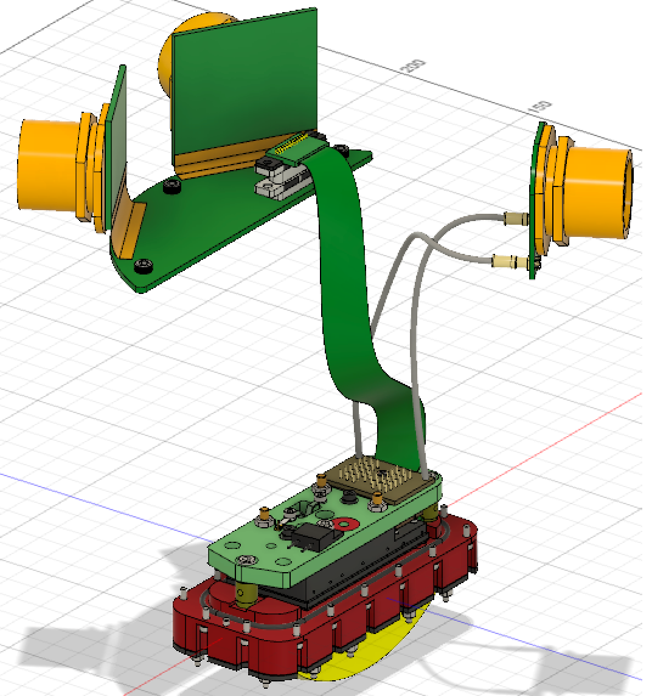}
   \end{tabular}
   \end{center}
   \caption[VIS Camera] 
   { \label{fig4} design of the CCD flex boards.}
   \end{figure}

\section{UV-VIS laboratory tests}  
The first test, to optimize the connections and minimize the noise, was done with a prototype PCB (Figure ~\ref{fig9}) and with a laboratory CCD housing (Figure ~\ref{fig10}). In these test we measured the CCD clocks, the bias and the video signals.

\subsection{Electronic tests}  
\begin{figure} [H]
   \begin{center}
   \begin{tabular}{c} 
   \includegraphics[height=7.3 cm]{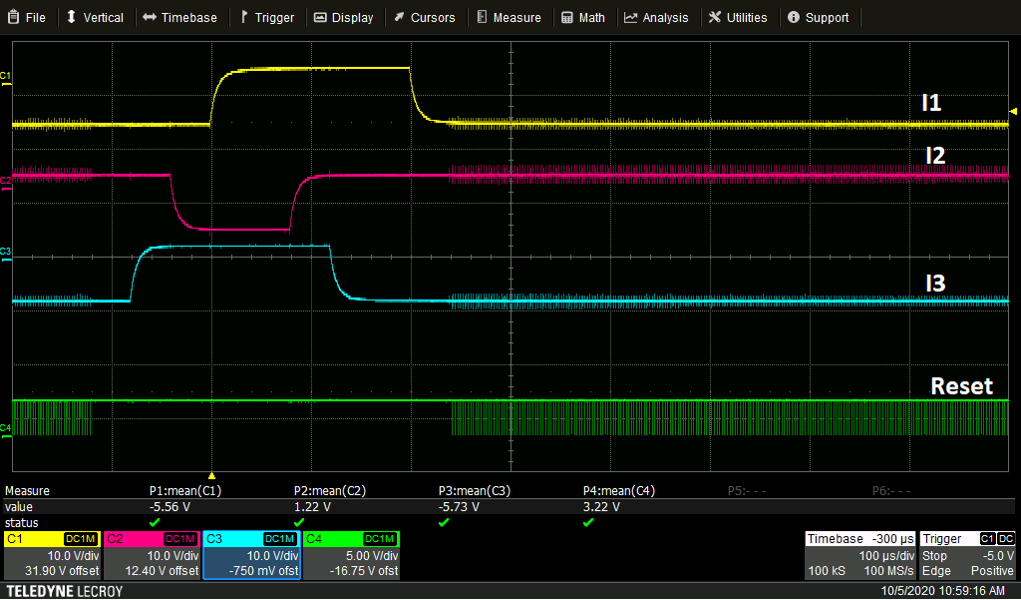}
   \end{tabular}
   \end{center}
   \caption[VIS Camera] 
   { \label{fig5} CCD vertical phases}
   \end{figure}

\begin{figure} [H]
   \begin{center}
   \begin{tabular}{c} 
   \includegraphics[height=7.3 cm]{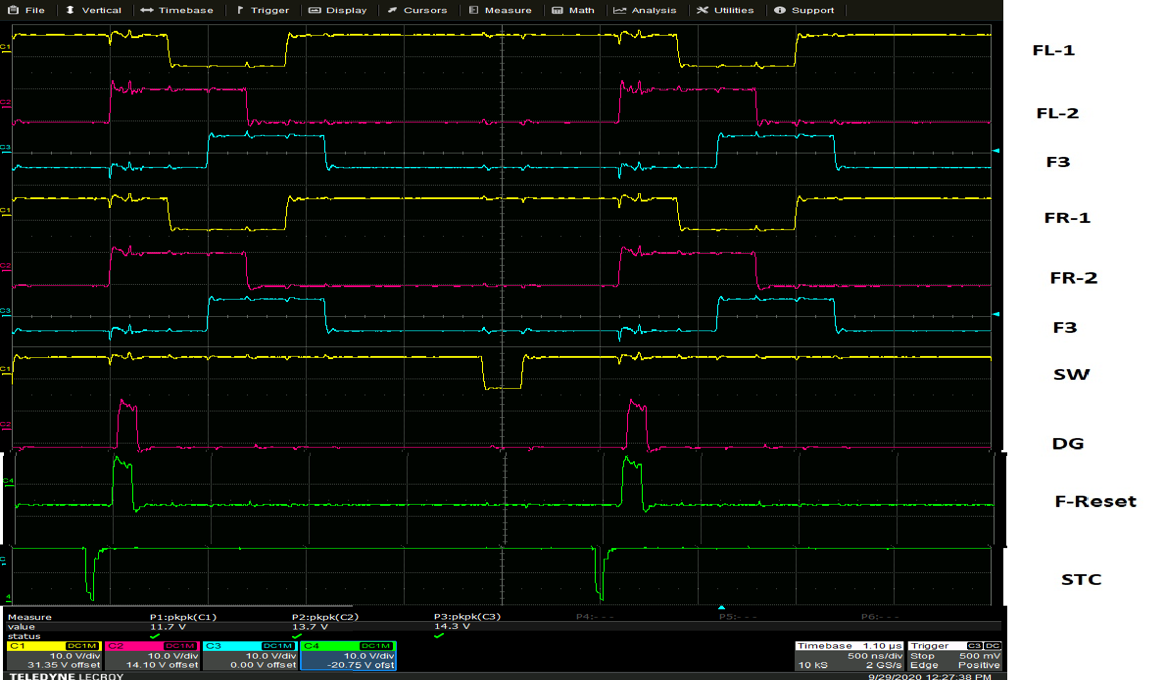}
   \end{tabular}
   \end{center}
   \caption[VIS Camera] 
   { \label{fig6} CCD horizontal phases and control timing.}
   \end{figure}

The measures of the clocks signals shown a correct temporization of the phases and the good quality of the signals on the Zero Insertion Force (ZIF) socket of the CCD. Figure ~\ref{fig5} and Figure ~\ref{fig6} shown the CCD phases measured in the CCD board (on the ZIF socket).

\begin{figure} [H]
   \begin{center}
   \begin{tabular}{c} 
   \includegraphics[height=8 cm]{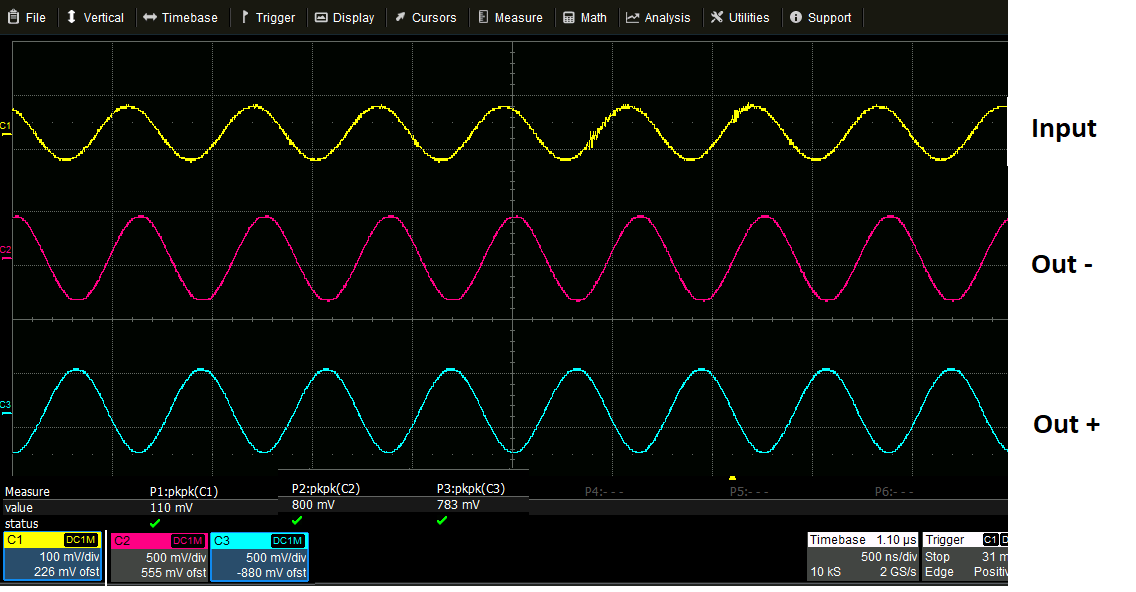}
   \end{tabular}
   \end{center}
   \caption[VIS Camera] 
   { \label{fig7} Preamplifier test with an external signal.}
   \end{figure}

The preamplifier was tested with a signal generator at different frequency to test the programmable bandwidth settings and the corresponding electronic gains (Figure ~\ref{fig7}).

\subsection{Detector tests}  
\begin{figure} [H]
   \begin{center}
   \begin{tabular}{c} 
   \includegraphics[height=8 cm]{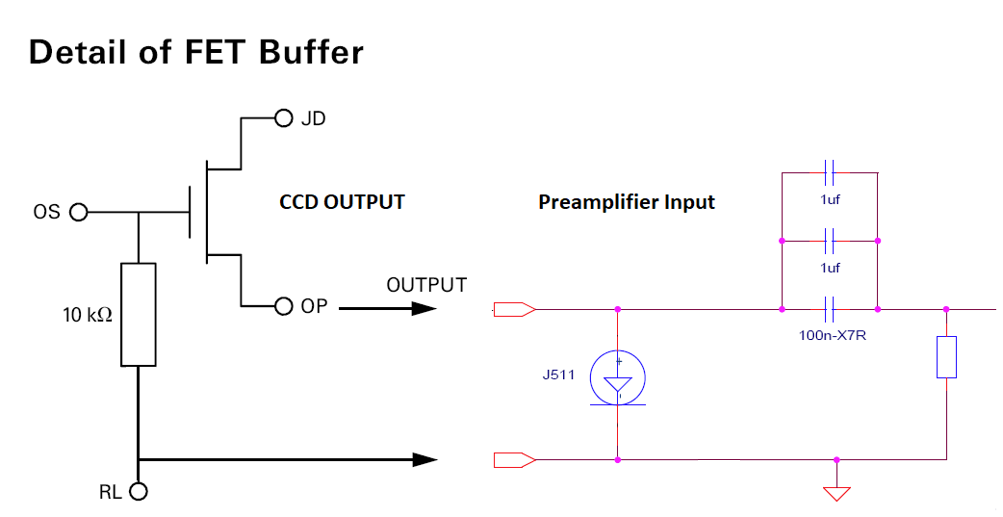}
   \end{tabular}
   \end{center}
   \caption[VIS Camera] 
   { \label{fig8} CCD output and preamplifier input.}
   \end{figure}

\begin{figure} [H]
   \begin{center}
   \begin{tabular}{c} 
   \includegraphics[height=8 cm]{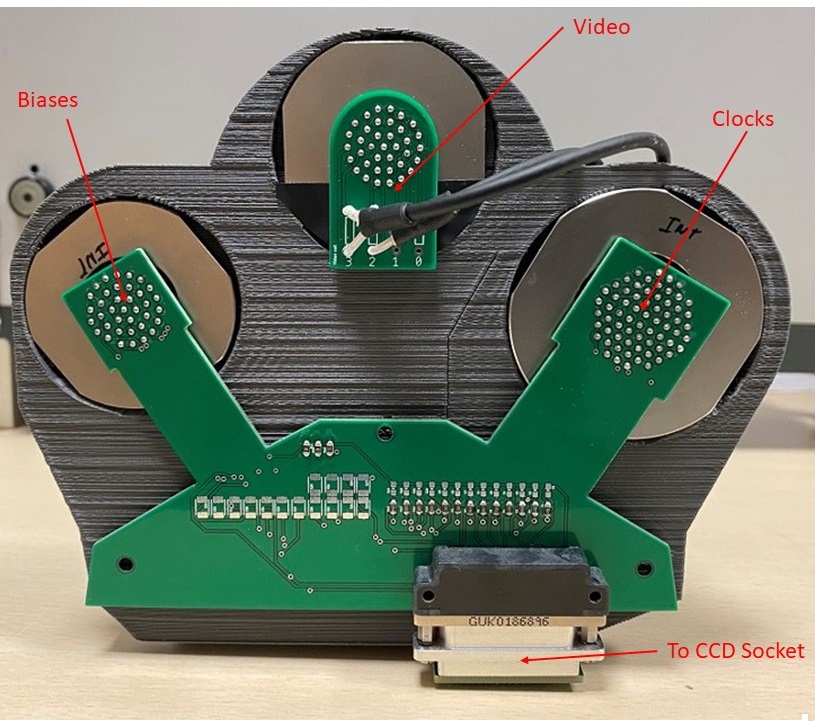}
   \end{tabular}
   \end{center}
   \caption[VIS Camera] 
   { \label{fig9} CCD board prototype.}
   \end{figure}

\begin{figure} [H]
   \begin{center}
   \begin{tabular}{c} 
   \includegraphics[height=4 cm]{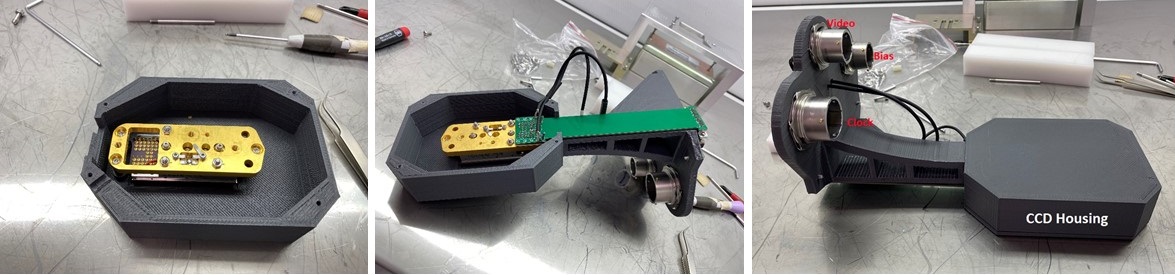}
   \end{tabular}
   \end{center}
   \caption[VIS Camera] 
   { \label{fig10} CCD test bench (for CCD warm tests}
   \end{figure}

The engineering CCD was tested with the test bench shown in Figure ~\ref{fig2}.
We chose the buffered CCD output OP and the preamplifier input as shown in Figure ~\ref{fig8}. 

\begin{figure} [H]
   \begin{center}
   \begin{tabular}{c} 
   \includegraphics[height=5 cm]{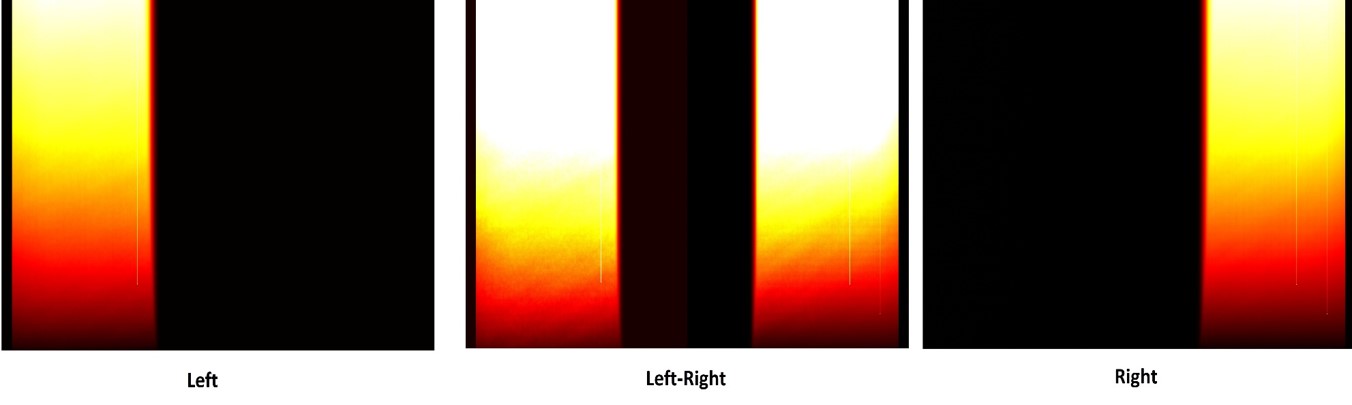}
   \end{tabular}
   \end{center}
   \caption[VIS Camera] 
   { \label{fig11} CCD images (room temperature).}
   \end{figure}

\begin{figure} [H]
   \begin{center}
   \begin{tabular}{c} 
   \includegraphics[height=4 cm]{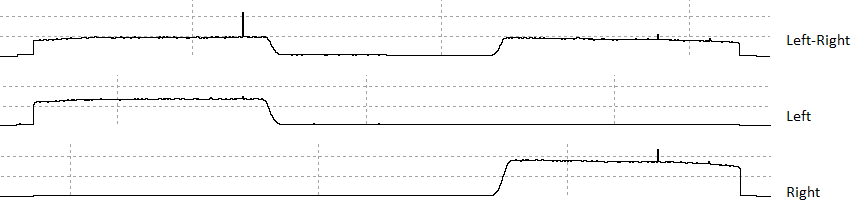}
   \end{tabular}
   \end{center}
   \caption[VIS Camera] 
   { \label{fig12} CCD vertical cuts.}
   \end{figure}

The tests have been done with an engineering CCD and at room temperature. In this condition is possible to verify if the acquisition system works as expected, because the CCD image has to show the overscan and the expected thermal signal. 
We have adjusted the video offsets and the gain and we acquired three images, one for each readout mode.
In Figure ~\ref{fig11} are shown the CCD images, acquired in the three readouts mode available (Left-Right, Left and Right). In these images the overscan is clearly visible and the images shows the thermic electrons. 
In this engineering CCD seems that the serial register is broken, because in the images and with more evidence in the horizontal cuts ( Figure ~\ref{fig12}) part of the CCD did not show the thermal signal.
   
The acquired images shown that the thermal signal increase with a gradient trend as expected (Figure ~\ref{fig12})

   More optimizations, at cryogenic temperature, will be done with the SOXS UV-VIS camera, the final flex PCB and the scientific CCD.
   \begin{figure} [H]
   \begin{center}
   \begin{tabular}{c} 
   \includegraphics[height=4 cm]{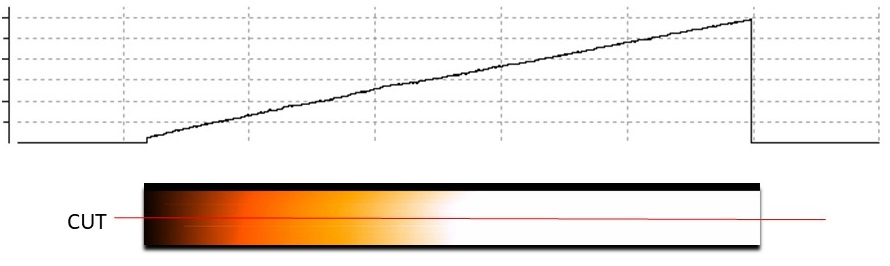}
   \end{tabular}
   \end{center}
   \caption[VIS Camera] 
   { \label{fig13} CCD horizontal cuts.}
   \end{figure}   
   
 \subsection{Vacuum tests}  
The preliminary vacuum test was done with the dummy CCD and without the CCD wiring. Leak test are foreseen for mid December 2020.
   
\subsection{Cryogenic tests}  
The cryogenic test was done with satisfactory results. The CFC shown a low LN2 consumption in a 10 days continuous cooling (less than 0.5 l/hr) and the equilibrium CCD temperature reach the expected values.

\subsection{Conclusions}
The tests on the prototype PCBs with the engineering CCD shown that the acquisition system works as expected. The next step will be the final test at cryogenic temperature with the scientific CCD in order to optimize the CCD settings and reach the optimum performance in term of thermic stability of the detector, minimum readout noise and data throughput.    

\acknowledgments 
 
A special acknowledgement to the European Southern Observatory for the support provided and for the availability to share its knowledge and to allow the use of the ESO laboratories in Garching.


 


\bibliography{VIS} 
\bibliographystyle{spiebib} 
\end{document}